\documentclass[twocolumn]{aastex631}

\newcommand{\gammaray}{$\gamma$-ray }
\newcommand{\Fermilat}{{\it Fermi}-LAT }

\usepackage{amsmath}

\shorttitle{Emission-line and gamma-ray dominance correlation in 3C\,279}
\shortauthors{Dmytriiev et al.}
\graphicspath{{./}{figures/}}

\begin{document}

\title{Correlation between emission-line luminosity and gamma-ray dominance in the blazar 3C\,279}

\correspondingauthor{A.~Dmytriiev}

\author{A.~Dmytriiev}
\affiliation{Centre for Space Research\\
North-West University\\
Potchefstroom, 2520, South Africa}
\email{anton.dmytriiev@nwu.ac.za}

\author{M.~B\"{o}ttcher}
\affiliation{Centre for Space Research\\
North-West University\\
Potchefstroom, 2520, South Africa}

\author{T.O.~Machipi}
\affiliation{Centre for Space Research\\
North-West University\\
Potchefstroom, 2520, South Africa}








\begin{abstract}

Despite numerous studies, the origin of the \gammaray emission from blazars is still debated, in particular whether it is produced by leptonic or hadronic processes. In this study, we are testing the leptonic scenario for the Flat Spectrum Radio Quasar (FSRQ) 3C\,279, assuming that the \gammaray emission is generated by inverse Compton scattering of external target photons from Broad Line Region (IC-BLR scenario). For this purpose we use a 10-year data set of the source consisting of the optical spectroscopy data from the Steward Observatory blazar monitoring program and \Fermilat \gammaray data. We search for a possible correlation between the Compton dominance and the emission line luminosity using the discrete correlation function (DCF) analysis. As a result, we find no significant correlation between these two quantities at any time lag value, while the emission line luminosity displays a moderate correlation with the \gammaray flux at a zero time lag. We also reveal that the optical synchrotron continuum flux shows a pronounced correlation with the \gammaray flux, and therefore we interpret these results within the leptonic IC-BLR scenario where the Compton dominance variations are primarily induced by changes in the magnetic field, rather than in the emission line luminosity.
\end{abstract}



\section{Introduction} \label{sec:intro}

Blazars represent radio-loud Active Galactic Nuclei (AGN) with a relativistic jet pointing very closely to the Earth's direction. Their broad-band spectral energy distribution (SED) exhibits two distinct bumps, with the low-energy one spanning typically from radio frequencies up to optical/UV/X-rays, and the high-energy one, extending from X-rays into the High Energy (HE, $E > 100$~MeV) and sometimes even the Very High Energy (VHE, $E > 100$~GeV) \gammaray range. While it is well established that the low-energy component corresponds to synchrotron radiation produced by high-energy particles in the jet, the mechanism responsible for generation of the high-energy component, as well as the exact location of the \gammaray emitting region in the jet, remains unclear. Blazars are generally divided into two classes, BL Lacertae (BL Lac) objects and Flat Spectrum Radio Quasars (FSRQs), with the latter being more luminous overall, having their SED bumps peaking at lower frequencies, and with the electromagnetic power output dominated by the \gammaray band \citep[e.g.,][]{fossati1998}. FSRQs also show signatures of different radiation fields, such as broad emission lines in their optical spectra, etc., while BL Lac objects generally display featureless optical spectra.

Two competing scenarios for the origin of blazar \gammaray emission were proposed, namely a leptonic and hadronic one. In a leptonic scenario, the high-energy SED bump arises due to the inverse Compton scattering process, in which high-energy electrons in the jet upscatter soft photons to high energies. In case of BL Lac objects, the target photons are thought to be synchrotron photons produced by the relativistic electrons \citep[synchrotron self-Compton or SSC scenario, e.g.][]{maraschi1992}, while in the case of FSRQs the target photons are considered to be dominated by the external radiation fields (external Compton scenario) of the accretion disk \citep[e.g.][]{dermer1992} and/or broad line region (BLR) \citep[e.g.][]{sikora1994} and/or dusty torus \citep[e.g.][]{blazejowski2000}. The optical/UV emission of the BLR and the infrared radiation of the dusty torus represent the re-scattered and re-processed emission from the accretion disk. In the external Compton scenario, the distance of the emitting zone $r$ from the black hole determines the dominant target radiation field: the accretion disk for $r \lesssim 0.01$ pc, the BLR for $0.01 \lesssim r \lesssim 0.1$ pc, and the dusty torus for $r \gtrsim 0.1$ pc. In the alternative hadronic scenario, the \gammaray emission is generated by processes involving hadrons, in particular photo-hadronic interactions \citep[e.g.][]{mucke2003}, or synchrotron radiation of protons with extremely high energies \citep[e.g.][]{aharonian2000}. Finally, in an intermediate class of scenarios, lepto-hadronic ones, both leptonic and hadronic emission processes contribute significantly to the multi-wavelength and multi-messenger emission \citep[e.g.][]{cerruti2015}. 

3C\,279 \citep[$z = 0.536$;][]{marziani1996} is one of the best studied FSRQs, and was the first FSRQ detected by the EGRET (Energetic Gamma-Ray Experiment Telescope) instrument on board the Compton Gamma-Ray Observatory in the 30 MeV -- 5 GeV band \citep{hartman1992}. The source is also the first FSRQ detected in the VHE \gammaray band \citep{magiccollab2008}. 3C\,279 features strong variability from radio frequencies up to the VHE \gammaray range, and was actively studied across all wavebands \citep[e.g.][]{hayashida2012,larionov2020}. The broad-band SED of this object is characterized by the typical two-bump structure with the low-energy bump exhibiting its maximum in the infrared (IR) regime, and the high-energy bump peaking between 0.1 GeV and a few GeV \citep[e.g.][]{hayashida2012,hayashida2015}.

In the radio band, 3C\,279 displays a compact milliarcsecond core. Radio observations performed in 1998 -- 2001 \citep{jorstad2004,jorstad2005} showed apparent superluminal velocities in the range of 4 -- 20 c, and allowed a measurement of the bulk Lorentz factor of the emitting region, $\Gamma \sim 15$, as well as the angle between the jet and the line of sight, $\theta \sim 2^{\circ}$, which translates to a Doppler factor of $\delta \sim 24$.

In the optical domain, 3C\,279 shows strong broad emission lines. The estimated accretion disk luminosity of this source is in the range $L_{\text{D}} \sim (0.2 - 0.6) \times 10^{46}$ erg s$^{-1}$ \citep{pian1999,hayashida2015}. The mass of the central supermassive black hole is in the range $M_{\text{SMBH}} \sim (3 - 8) \times 10^8 \, M_{\odot}$, estimated via three independent methods \citep{woourry2002,gu2001,nilsson2009}. 3C\,279 also displays a quite significant degree of optical polarization with variations in the $U$-band as high as 45.5\% \citep{mead1990}. 

In the \gammaray range, 3C\,279 was continuously observed by the \textit{Fermi} Gamma-Ray Telescope since it started operating in 2008.
The object is highly variable in $\gamma$-rays. During the period December 2013 -- April 2014, 3C\,279 underwent a high state featuring a series of flaring events, with the highest level of \gammaray flux of $F(> 0.1 \text{ GeV}) \simeq 10^{-5}$ ph cm$^{-2}$ s$^{-1}$, and the shortest observed flux-doubling time-scale of only 2 h \citep{hayashida2015}. The source also showed a very hard \gammaray spectrum during this high state, with a spectral index $\alpha_{\gamma} = 1.7 \pm 0.1$, which is rarely observed in FSRQs, and it achieved a very high level of Compton ($\gamma$-ray) dominance, i.e.\ a high ratio of the \gammaray flux over the synchrotron flux, $F_{\gamma}/F_{\text{syn}} > 300$ \citep{hayashida2015}. On June 16, 2015, 3C\,279 showed a violent outburst with the historically highest peak \gammaray flux of $F(> 0.1 \text{ GeV}) = (3.6 \pm 0.2) \times 10^{-5}$ ph cm$^{-2}$ s$^{-1}$ translating into an equivalent isotropic \gammaray luminosity of $L_{\gamma} \sim 10^{49}$ erg s$^{-1}$, and extremely short flux-doubling time-scales of $\lesssim 5$ min \citep{ackermannGeVflare}. During the period January -- June 2018, the source displayed a range of flares characterized by a complex ``peak-in-peak'' variability pattern with the fast flux variations occurring at time-scales of minutes seen on top of a slower envelope varying on a $\sim 1$ d time-scale \citep{shuklamannheim2020}. Based on the flux behavior, the authors suggest that these flares are powered by relativistic magnetic reconnection.

A number of studies and modeling efforts were carried out to understand the origin of the \gammaray emission from 3C\,279 and determine the location of the emitting region. Different multi-band data sets collected during various stages of activity are satisfactorily described with leptonic \citep[e.g.][]{boettcher2013,dermer2014,paliya2015}, hadronic \citep[e.g.][]{boettcher2009,boettcher2013,petropoulou2017} and lepto-hadronic models \citep[e.g.][]{paliya2018}. For example, \cite{dermer2014} consider a leptonic model based on an external Compton scenario with equipartition between the energy density of particles and the magnetic field, and find that the model generally provides a reasonable description of a selection of multi-epoch SEDs of 3C\,279, although it underestimates the level of multi-GeV \gammaray emission. In another study by \cite{paliya2015}, the authors show that the March -- April 2014 flare of the source can be well fitted with a simple one-zone leptonic scenario involving an increase in the bulk Lorentz factor, with both BLR and dusty torus fields necessary to reproduce the measured \gammaray spectrum, therefore placing the emitting zone close to the BLR. Further on, based on the \Fermilat data of the source, \cite{acharyya2021} conclude that its \gammaray emission is consistent with a BLR origin. To the contrary, in an attempt to explain the origin of the VHE \gammaray emission detected by MAGIC in 2006, \cite{boettcher2009} disfavor the external Compton process on BLR seed photons, since such a scenario requires unrealistic physical parameters and yields much steeper $\gamma$-ray spectra than observed, whereas a hadronic description with photo-hadron interactions gives quite reasonable results. Given the lack of correlation between the optical band and the VHE $\gamma$-rays, the authors suggest a multi-zone scenario in which these two emissions originate from different locations in the jet. In addition, if the emitting zone was located in the vicinity of the BLR, the VHE $\gamma$-rays would be severely attenuated due to $\gamma$-$\gamma$ absorption on the BLR photon field, which contradicts the VHE detection by MAGIC. To remedy this problem, \cite{bottcherels} propose another type of multi-zone model, in which the VHE and GeV emissions are not generated co-spatially, with the GeV $\gamma$-rays being produced close to BLR at sub-parsec distances, while the VHE emission originates far away from the BLR at multi-parsec distances.  

Exploring correlations between different wavebands is another powerful tool to disentangle emission scenarios, in particular, within a leptonic scenario, one expects presence of a correlation between the flux variations in the optical and \gammaray regimes as those emissions are produced by electrons of similar energies. So far however, various studies of such correlations indicate controversial results. Based on a multi-frequency data set including the first two years of \Fermilat data (2008 -- 2010), \cite{hayashida2012} discover a prominent correlation between the optical and \gammaray bands, with $\gamma$-ray flux changes leading the ones in the optical domain by $\sim 10$ days. An opposite behavior was reported by \cite{hayashida2015} for the December 2013 -- April 2014 \gammaray outburst of the object, where the optical flux displayed no significant correlation with the one in $\gamma$-rays, suggesting that the optical emission might be produced in multiple regions along the jet. The authors also find it challenging to explain the observed MWL emission properties, in particular the very hard \gammaray spectrum, with a one-zone leptonic model. In another study by \cite{larionov2020}, based on a decade-long multi-band data set of the object, the authors conclude that while X-ray and \gammaray light curves display a clear correlation, the relationship between the optical and \gammaray fluxes is rather complex with the exact dependence varying from one activity state to another.

In this work, we aim at testing the leptonic scenario for \gammaray emission production, specifically a hypothesis that the GeV emission from the source is generated via inverse Compton upscattering of soft photons from the BLR (IC-BLR scenario) by high energy electrons residing in a compact region of the jet (a blob) moving relativistically down the jet with a bulk Lorentz factor $\Gamma$. Under this scenario, one expects to observe a correlation between the luminosity of emission lines (originating in the BLR) and the Compton Dominance parameter. The Compton Dominance (CD) is defined as the ratio between the inverse Compton and the synchrotron flux, $CD = F_{\mathrm{IC}}/F_{\mathrm{syn}}$, with $F_{\mathrm{syn/IC}} = (\nu F_{\nu})_{\mathrm{syn/IC}}$. This quantity is proportional to $(N_e U_{\mathrm{rad}}^{\prime})/(N_e U'_{\mathrm{B}}) \propto U_{\mathrm{rad}} \Gamma^2/B^2$, with $U_{\mathrm{rad}}$ being the energy density of the target field (in the AGN rest frame), which under our hypothesis is the BLR emission, so that $U_{\mathrm{rad}}$ is proportional to the flux of the emission lines as measured by a distant observer. Exploring a correlation between the emission line luminosity and the Compton dominance, rather than just between the optical and \gammaray emissions, not only allows us to test the hypothesis that same electron population is responsible for generation of synchrotron and \gammaray emissions (leptonic scenario), but also the IC-BLR assumption.

The paper is organized as follows. In Section~\ref{sec:data} we present the 10-year-long data set that was used in our study, as well as respective analysis methods to arrive at the measurement of emission line luminosity and Compton dominance versus time. In Section~\ref{sec:correlation} the correlation analyses among various quantities are performed, with the direct (simultaneous) correlation being explored first, and the discrete correlation function (DCF) calculated in the next step. Section~\ref{sec:discussion} provides a detailed discussion of the obtained results, and Section~\ref{sec:summary} a summary of the key results and an outlook.

\section{Data} \label{sec:data}

The optical emission line flux and the synchrotron continuum flux, as a function of time, are determined by using the optical spectroscopy data from the Steward Observatory blazar monitoring program. To obtain a measurement of the \gammaray flux of the source as a function of time, i.e.\ a \gammaray light curve, we use the \gammaray data from the \Fermilat instrument. We estimate the Compton dominance as the ratio of the \gammaray energy flux of the source in the energy range of 0.1 -- 100 GeV, to the optical synchrotron $\lambda F_{\lambda}$ flux of the source measured at the wavelength of 6000 \AA. The 0.1 -- 100 GeV range covers most of the high-energy SED bump of 3C\,279 (including the peak at $\sim$1 GeV), that arises presumably due to the external Compton (EC) mechanism. We choose to measure the synchrotron continuum at 6000 \AA, because the SED around this wavelength is dominated by non-thermal synchrotron continuum with only a negligible contribution of thermal emission from the accretion disk \citep[e.g.,][]{rajput2020} which, if in a high state, or in case of a low level of the continuum, starts to emerge typically at wavelengths shorter than 5000 \AA. Although the low-energy bump of 3C\,279 SED is peaking in the infrared rather than in the optical regime, the quantity $\lambda F_{\lambda}(\lambda = 6000 \text{ \AA})$ should be commensurate to (or order of magnitude of) the same quantity measured at the peak (and being roughly proportional to it, assuming the spectral index variations are sub-dominant). The ratio we calculate therefore provides a reasonable estimate of (and is proportional to) the Compton dominance evaluated in a standard way.

\subsection{Steward Observatory} \label{subsec:Steward}

\subsubsection{Optical data} 

For 10 years (2008 - 2018), the Steward Observatory of the University of Arizona performed regular monitoring of a selection of \Fermilat blazars in the optical band, with the aim to contribute to the multi-wavelength coverage of blazars. The Steward blazar monitoring program\footnote{\url{http://james.as.arizona.edu/~psmith/Fermi/}} provides data of optical linear polarization (polarization degree and polarization spectra), as well as of optical spectra and photometry. We use the optical spectra of 3C\,279 from this program to obtain measurements of the emission line luminosity and of the synchrotron continuum flux, both as a functions of time. In our analysis, we only select spectra for which the normalization of the spectral flux density has been properly scaled, so that this normalization agrees with the $V$-band magnitude obtained from the differential spectrophotometry. The available data includes 504 spectra, stretching a wavelength domain of 4000 -- 7550 \AA, and measured over the time period of 2008 November 24 -- 2018 July 7 (with a somewhat irregular time sampling).

The optical spectra of the source typically represent a combination of a continuum with a power law shape (roughly) and a number of emission lines. A change in the spectral shape (compared to the red part of the spectrum) is seen at blue wavelengths, characterized either by a moderate attenuation occurring due to Galactic extinction, or by an opposite effect -- a significant pile-up at wavelengths below 5000 \AA, appearing due to a significant contribution from the accretion disk. On the other (red) side of the spectrum, an $O_2$ B-band absorption feature is present in the majority of the spectra at about 6980 \AA. The continuum emission component is produced in the jet, whereas the emission lines arise in the BLR. The most prominent emission line in the spectra is found to be Mg II ($\lambda = 2798 \AA$ in the source's frame, and redshifted to $\lambda = 4298 \AA$ in the observer's frame), while the other emission lines appear to be much weaker and are difficult to detect clearly over the continuum, and therefore we have not considered them in our analysis.

\subsubsection{Fitting procedure}

A Python module \texttt{LMFIT} is used to perform fitting and derive relevant model parameters for each spectrum and an associated observational date. In order to measure the synchrotron continuum level, we fit each optical spectrum with a power law function in a restricted wavelength range of 5000 - 6500 \AA, where we avoid a possible contamination by the accretion disk, as well as the strong $O_2$ absorption line. The synchrotron continuum emission is described by a power law of the form

\begin{equation} \label{eq:powerlaw}
    F_{\lambda,\mathrm{SC}} = A_{\mathrm{sc}} (\lambda/\lambda_1)^{-p_1} 
\end{equation}

with $A_{\mathrm{sc}}$ being the normalization, $p_1$ the spectral index of the synchrotron continuum in the range 5000 -- 6500 \AA, and $\lambda_1 = (\lambda_{\mathrm{min},1} \lambda_{\mathrm{max},1})^{1/2} = (5000 \times 6500)^{1/2} \text{\AA} \approx 5700$ \AA\ is the pivot wavelength.

Using the best-fit model (for a given spectrum), we evaluate the quantity $\lambda F_{\lambda}$ at the wavelength of 6000 \AA \, and treat it as a measurement of the synchrotron continuum flux, $F_{\mathrm{syn}} \sim (\lambda F_{\lambda,SC})|_{\lambda = 6000 \text{ \AA}}$. We also calculate the uncertainty for $F_{\mathrm{syn}}$, which is controlled by the uncertainties of the continuum normalization $\Delta A_{\mathrm{sc}}$ and of the spectral index $\Delta p_1$:

\begin{multline} \label{eq:uncertsynflux}
    (\Delta F_{\mathrm{syn}})^2 = \left(\tfrac{\partial F_{\mathrm{syn}}(A_{\mathrm{sc}},p_1)}{\partial A_{\mathrm{sc}}} \, \Delta A_{\mathrm{sc}}\right)^2 + \left(\tfrac{\partial F_{\mathrm{syn}}(A_{\mathrm{sc}},p_1)}{\partial p_1} \, \Delta p_1\right)^2 \\ + \, 2 \, \tfrac{\partial F_{\mathrm{syn}}(A_{\mathrm{sc}},p_1)}{\partial A_{\mathrm{sc}}} \tfrac{\partial F_{\mathrm{syn}}(A_{\mathrm{sc}},p_1)}{\partial p_1} \, \text{cov}(A_{\mathrm{sc}},p_1)
\end{multline}

with $F_{\mathrm{syn}}(A_{\mathrm{sc}},p_1) = \lambda F_{\lambda,\mathrm{SC}}$, and $\text{cov}(A_{\mathrm{sc}},p_1)$ being the covariance between the normalization and spectral index, which is the element of the covariance matrix computed in the parameter optimization procedure.

To infer the Mg II emission line flux, we fit the optical spectra with a sum of a power law and a Gaussian in a relatively narrow wavelength domain around the emission line (4150 -- 5000 \AA), in order to probe the continuum locally (rather than globally), as well as to exclude the typically noisy extreme blue end of the spectrum. Here, we are not performing a detailed physical modeling of the complex continuum shape at blue wavelengths, and instead, simply approximate the local continuum within the considered wavelength range as a power law, which appears to be reasonable, given the narrow width of this range. In such a way, we obtain the flux of the emission line above a given local continuum, thus significantly reducing a potential bias in the line flux measurement. The composite model to fit the Mg II emission line with the local continuum is

\begin{equation} \label{eq:gaussianpluscontmodel}
    F_{\lambda,\mathrm{LC}} = A_{\mathrm{c}} (\lambda/\lambda_2)^{-p_2} \, + \, \tfrac{A_{\mathrm{g}}}{\sigma_{\mathrm{g}} \sqrt{2 \pi}} \text{exp} \left(- \tfrac{(\lambda - \lambda_{\mu})^2}{2 \sigma_{\mathrm{g}}^2} \right)
\end{equation}

where $A_{\mathrm{c}}$ is the normalization, and $p_2$ the spectral index of the local continuum in the range 4150 -- 5000 \AA, $\lambda_2 = (\lambda_{\mathrm{min,2}} \lambda_{\mathrm{max,2}})^{1/2} = (4150 \times 5000)^{1/2} \text{\AA} \approx 4555$ \AA\ is the pivot wavelength, $A_{\mathrm{g}}$ is the normalization of the Gaussian equal to the total flux contained in the emission line (in erg cm$^{-2}$ s$^{-1}$), $\sigma_{\mathrm{g}}$ is the line width in \AA, and $\lambda_{\mu}$ is the central wavelength, fixed to $\lambda_{\mu} = 4298 \AA$.

For a given optical spectrum, the best-fit (local) continuum parameters in this model often appear to be somewhat different from the synchrotron continuum (Eq.~\ref{eq:powerlaw}) parameters, due to the contribution of the accretion disk or due to the effect of Galactic extinction. The two power law continua can be joined in one function describing the global continuum either as a broken power law (in case of significant accretion disk contribution) or a power law with an exponential suppression at short wavelengths (in case of a prominent Galactic extinction effect). However, in this work, we choose to fit the Mg II emission line and the synchrotron continuum separately/independently in their respective narrow spectral ranges, rather than performing a global fit / physical modeling of the spectrum in the entire wavelength range (in which we are not interested here), as this approach appears to be sufficient for our objectives, and at the same time allows for accurate measurement of the fluxes. It is worth to note that the retrieved line fluxes are still not corrected for Galactic extinction, however since all the values are affected in the same way, the quantity we measure is always proportional to the true emission line flux with the same factor. Therefore, in this analysis one can disregard the effect of Galactic extinction.

For a rather important share of the studied spectra, the Mg II line is very difficult or impossible to clearly identify due to the noise and/or very high level of the continuum. In such a situation, the fitting routine identifies continuum fluctuations as a very broad or very narrow emission line. We judge the detection of the Mg II emission line by eye for each spectrum, requiring the line to be clearly visible above the continuum variations, as well as exclude spectra in which the line width $\sigma_{\mathrm{g}}$ is outside the range 21 -- 130 \AA, which corresponds to a range of typical velocities of BLR clouds, $\sim$ 1500 -- 9000 km s$^{-1}$. The resulting sample of optical spectra where the Mg II emission line was clearly identified, and its flux was reliably measured, comprises 169 spectra. At the same time, the synchrotron continuum flux $F_{\mathrm{syn}}$ with its uncertainty is measured for every spectrum from the entire collection of 504 optical spectra.

\begin{figure*}[t]
\includegraphics[width=0.8\textwidth]{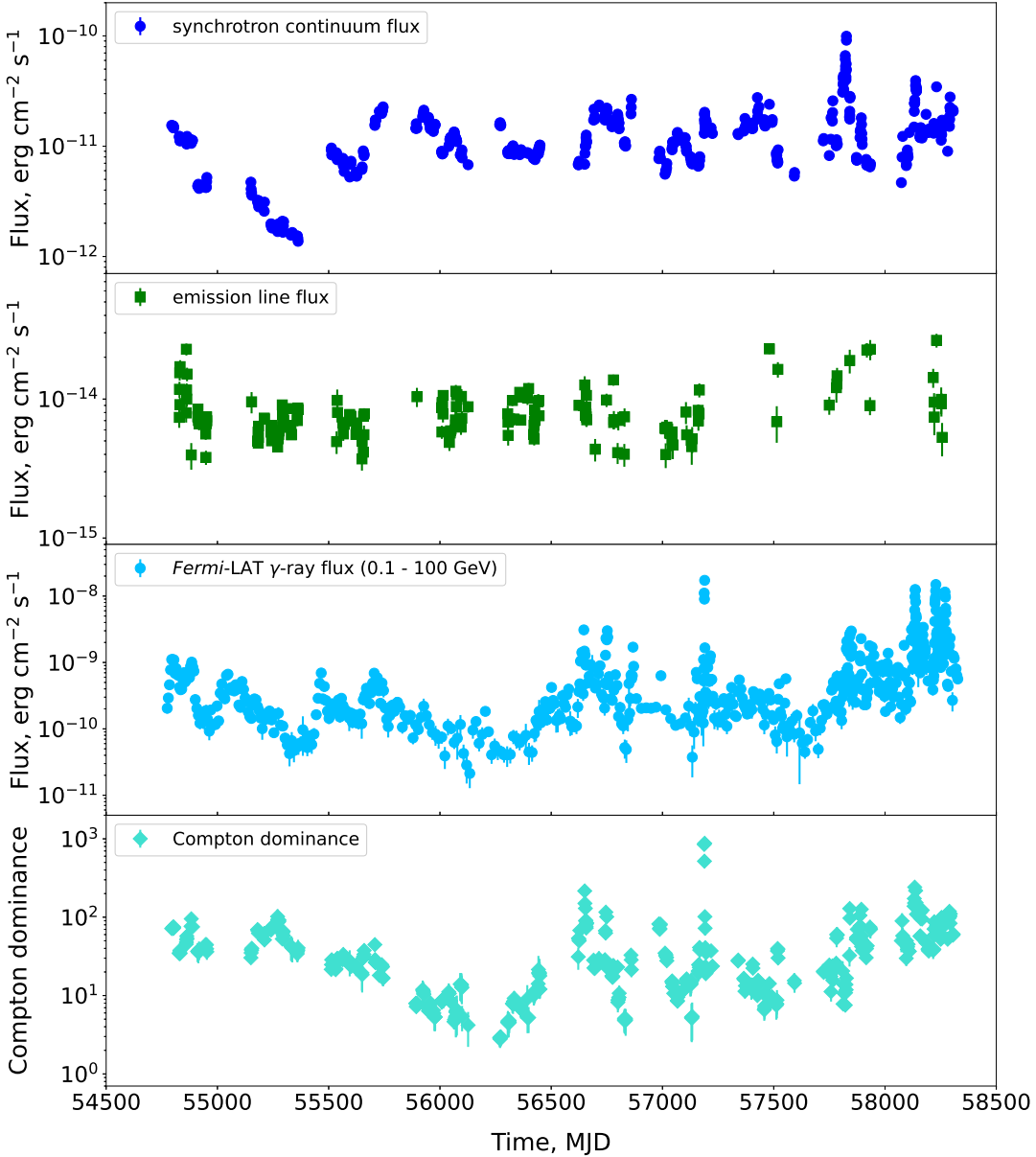}
\centering
\caption{First panel: continuum (optical synchrotron) $\lambda F_{\lambda}$ flux of 3C\,279 measured at a wavelength $\lambda = 6000$ \AA\ as a function of time (full sample of 504 spectra). Second panel: Mg II emission line flux $A_{\mathrm{g}}$ in the optical spectrum of 3C\,279 as a function of time (sub-sample of 169 spectra). Third panel: \Fermilat \gammaray light curve (energy flux vs time) of 3C\,279 in the energy range 0.1 -- 100 GeV (light-blue data points). Fourth panel: The Compton dominance of 3C\,279 as a function of time (turquoise data points).}
\label{fig:fourlightcurves}
\end{figure*}

As a result, we obtain the two key quantities necessary for our further analysis, namely the synchrotron SED flux $F_{\mathrm{syn}}$ and Mg II emission line flux (in erg cm$^{-2}$ s$^{-1}$), both as a function of time, for the period of 10 years (2008 -- 2018). The light curves are depicted in Fig.~\ref{fig:fourlightcurves} (first and second panels, respectively).

\begin{figure}[t]
\includegraphics[width=0.45\textwidth]{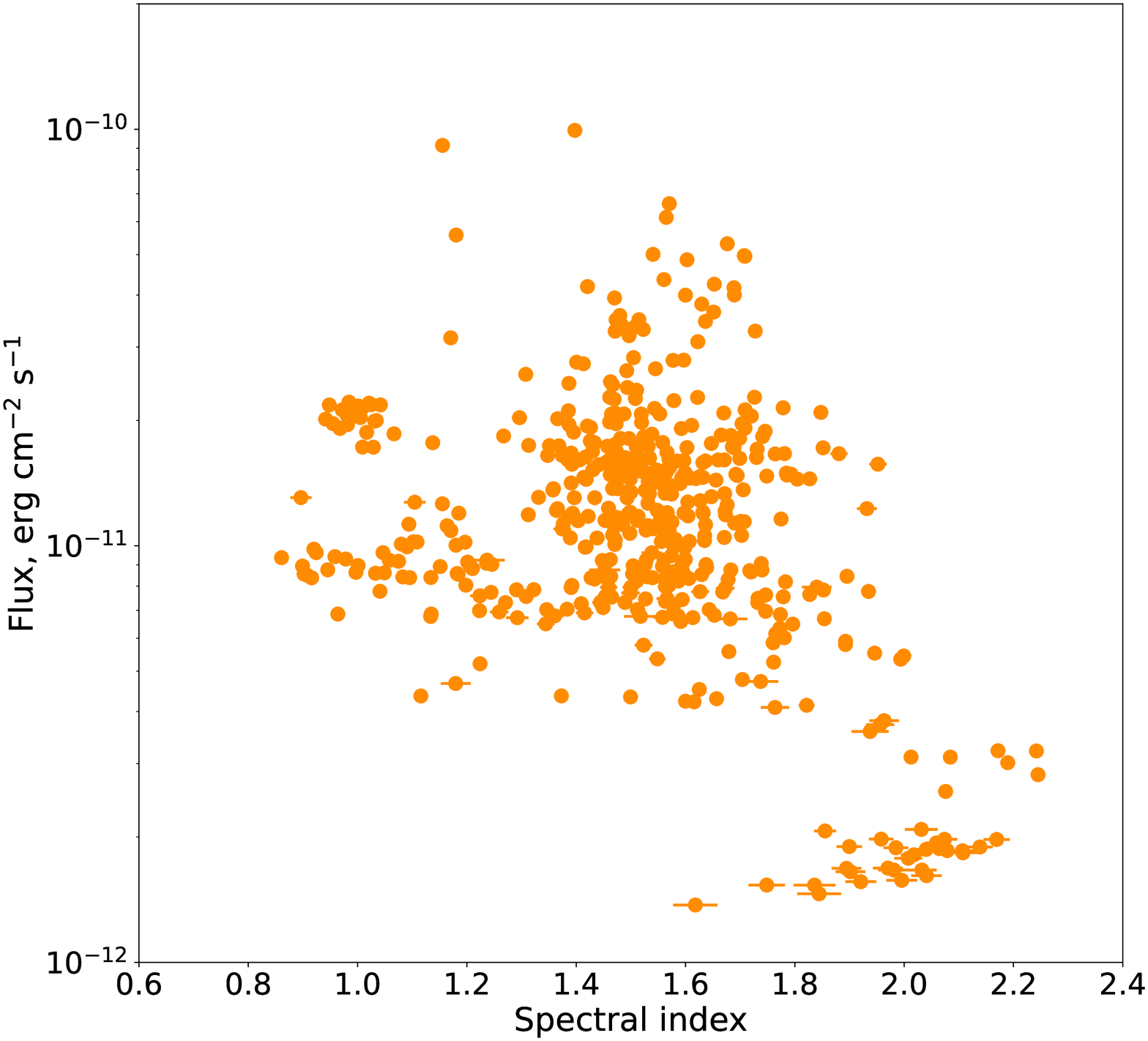}
\centering
\caption{Continuum (optical synchrotron) $\lambda F_{\lambda}$ flux of 3C\,279 measured at a wavelength $\lambda = 6000$ \AA\ versus spectral index of the $F_{\nu}$ spectrum (equal to $2-p_1$) measured in the domain 5000 - 6500 \AA\ (full sample of 504 spectra).}
\label{fig:scattersynflvsindex}
\end{figure}

As a by-product of our analysis, we also produce a color-magnitude (hysteresis) diagram with the synchrotron continuum ($\lambda F_{\lambda} = \nu F_{\nu}$) flux plotted against the spectral index $\alpha_1$ of the $F_{\nu}$ spectrum ($F_{\nu} \propto \nu^{-\alpha_1}$) measured in the range 5000 - 6500 \AA. The index of the $F_{\nu}$ spectrum is related to the one of $F_{\lambda}$ as $\alpha_1 = 2 - p_1$. The relevant plot is displayed in Fig.~\ref{fig:scattersynflvsindex}. One can see no clear overall correlation, but certain clusters of data points seem to suggest softer slopes for low continuum levels, and harder slopes for medium-to-high continuum fluxes, however with a number of outliers.

\subsection{Fermi-LAT} \label{subsec:Fermi}

We analyze \Fermilat 3C\,279 data spanning 10 years, from 2008 till 2018. The analysis is performed using the \textit{Fermipy} package (fermipy version 1.0.1, ScienceTools version 2.0.8), following the maximum likelihood optimization approach. We choose events from the time range of 2008 November 1, 00:00:00 -- 2018 July 31, 23:59:59 UTC (a mission elapsed time (MET) range of 247190402 -- 554774404), which corresponds to the time range of the available optical data (see the next section), with some margin at both ends. We select events of the ``SOURCE'' class (\texttt{evclass=128}), from which we further select FRONT+BACK converting events within all PSF and energy subclasses (\texttt{evtype=3}), from a circular region of interest (ROI) with a radius of 15$^{\circ}$ centered around the target source, and from the energy range of 0.1 -- 100 GeV, as well as apply a zenith angle cut of $z < 90^{\circ}$. A standard filter expression 'DATA\_QUAL $> 0$ \&\& LAT\_CONFIG == 1' is used in our analysis configuration. We use two background components: Galactic diffuse background, modeled with the most recent \textit{gll\_iem\_v07} spatial and spectral template, and isotropic diffuse background, taken into account using the newest currently available spectral template, \textit{iso\_P8R3\_SOURCE\_V3\_v1}. The \texttt{P8R3\_SOURCE\_V3} instrument response functions (IRFs) are used in our analysis, in coherence with the background templates versions. The point sources that lie within the ROI are added to XML model from the 4FGL catalog. After running the ``optimize'' method of \textit{Fermipy}, we delete from the ROI model all points sources with TS $< 9$ (equivalent to significance of detection below $3\sigma$ level) and with number of predicted counts $n_{\mathrm{pred}} < 3$, therefore leaving only relatively prominent ones. After that, for the remaining point sources, we free the normalizations of all sources having TS $> 10$, as well as of those within $3^{\circ}$ of the ROI center. We also free all the parameters of the Galactic diffuse and isotropic diffuse backgrounds, as well as of the source of interest, 3C\,279.

We then apply the ``lightcurve'' method of \textit{Fermipy} to the analyzed data to calculate the light curve of 3C\,279. An adaptive time binning is chosen for the light curve in order to get a decent time resolution for high-flux states, and avoid large uncertainties (or upper limits) for low-flux states. We employed a significantly simplified version of the technique used by \cite{larionov2017} and \cite{larionov2020}, where instead of iterative variation of the time bin width around a certain MJD to achieve an optimal TS, we first extract a \Fermilat light curve with a fixed and relatively sparse binning in order to identify different activity states of the source, and then form a time grid with a variable bin width, based on the flux level during a given epoch. Following this approach, we initially compute the \Fermilat light curve with a fixed 1-week (7 d) time binning. Next, we process the obtained light curve to find different flux levels, and apply a 1-day time binning to the brightest flares with a peak exceeding the arbitrarily chosen threshold $F_{\gamma} > 8 \times 10^{-10}$ erg cm$^{-2}$ s$^{-1}$, a 3-day binning to periods with flux variations $3 \times 10^{-10} < F_{\gamma} \leq 8 \times 10^{-10}$ erg cm$^{-2}$ s$^{-1}$, a 7-day binning for periods with $10^{-10} < F_{\gamma} \leq 3 \times 10^{-10}$ erg cm$^{-2}$ s$^{-1}$, a 10-day binning when $8 \times 10^{-11} < F_{\gamma} \leq 10^{-10}$ erg cm$^{-2}$ s$^{-1}$, and a 14-day binning when $F_{\gamma} \leq 8 \times 10^{-11}$ erg cm$^{-2}$ s$^{-1}$. We then extract the \Fermilat light curve on the obtained adaptive-bin time grid, and as a result, get the energy flux measurement (in erg cm$^{-2}$ s$^{-1}$) for 3C\,279 in each time bin in the energy range 0.1 -- 100 GeV. We have verified that our adaptive time binning approach is indeed optimal and yields satisfactory results in terms of the reliable source detection in each time bin and optimal (not too high) flux uncertainty values. The resulting long-term \Fermilat light curve of the source is shown in Fig.~\ref{fig:fourlightcurves} (third panel).

\section{Correlation analyses} \label{sec:correlation}

\subsection{Calculation of Compton Dominance}

We now proceed to the computation of the Compton Dominance (CD) as a function of time. The CD is defined as the ratio of the peak inverse Compton (IC) $\nu F_{\nu}$ flux to the synchrotron one. As already discussed in Section~\ref{sec:data}, we use the \Fermilat \gammaray light curve we obtained in Section~\ref{subsec:Fermi} as the measure of the IC $\nu F_{\nu}$ flux, and the light curve of the optical synchrotron $\lambda F_{\lambda}$ flux (at $\lambda = 6000$ \AA) we computed in Section~\ref{subsec:Steward} as the quantity approximately proportional to the $\nu F_{\nu}$ synchrotron flux. Since, in contrast to the \gammaray data, the optical measurements have a rather irregular sampling, for the time binning of the CD we use the one of the synchrotron light curve, and for each of its data points we determine the corresponding \gammaray flux from the \Fermilat light curve, and evaluate the CD. Therefore, for each observational ID of the optical observations, we calculate the CD. We also compute the uncertainties on the CD values using the standard rule for an uncertainty of a ratio. The resulting CD light curve is depicted in Fig.~\ref{fig:fourlightcurves} (fourth panel).

\subsection{Direct simultaneous correlation}

\begin{figure}[b]
\includegraphics[width=0.45\textwidth]{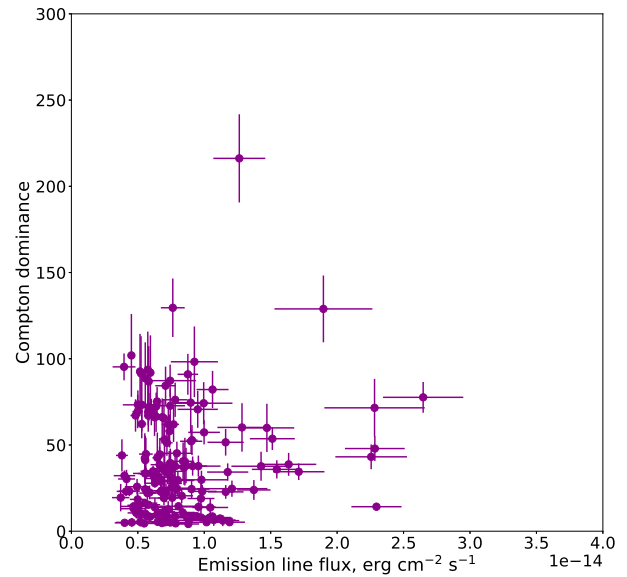}
\centering
\caption{The Compton dominance versus the Mg II emission line flux for simultaneous measurements of these quantities. Both axes are represented in a linear scale to facilitate the identification of a possible linear correlation.}
\label{fig:cdvslineflux}
\end{figure}

To test for a possible correlation between the emission line luminosity and the CD, we produce a scatter plot of simultaneous (the same observational ID) measurements of emission line flux and CD. The plot is displayed in Fig.~\ref{fig:cdvslineflux}. One can see essentially no correlation between the simultaneous observations of the two quantities. We then check a possibility that the two variables might show correlation with a time lag between them.        

\subsection{Discrete correlation function analysis}

\begin{figure*}[t]
\includegraphics[width=0.83\textwidth]{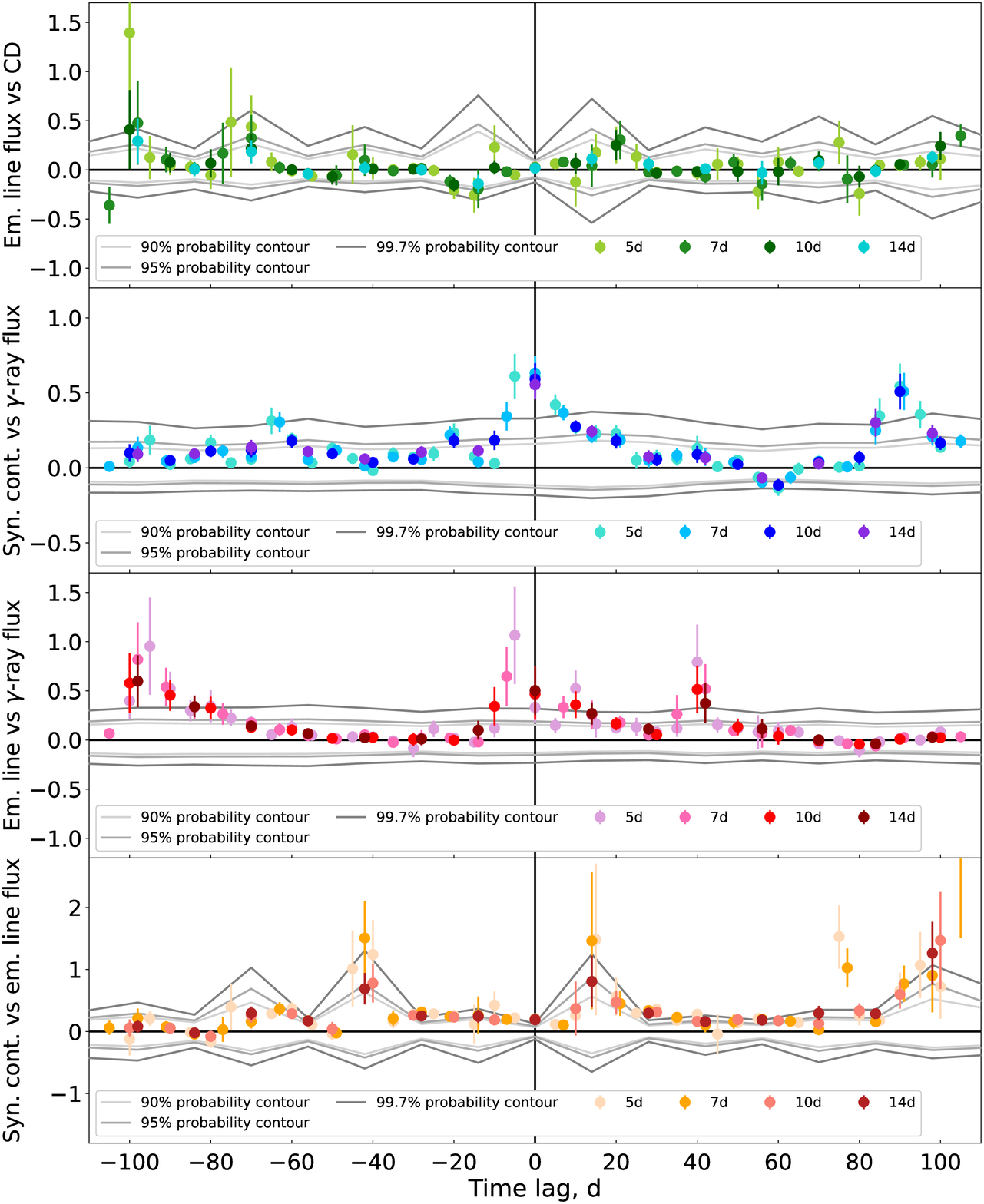}
\centering
\caption{DCF as a function of time lag with different time lag binnings (5, 7, 10 and 14 day bin widths) for (1) emission line flux versus Compton dominance (\textit{top}), (2) optical synchrotron continuum versus \gammaray flux (\textit{second from the top}), (3) emission line flux versus \gammaray flux (\textit{third from the top}), and (4) optical synchrotron continuum versus the emission line flux (\textit{bottom}). Different colors represent different binnings in the time lag space (5, 7, 10 and 14 day time bin widths). The gray contours, from lightest to darkest shades, indicate 90\%, 95\% and 99.7\% ($3\sigma$) probability contours, respectively.}
\label{fig:dcflfcdscgr}
\end{figure*}

The standard IC-BLR scenario predicts the presence of a time lag between the production of emission lines and an increase of the \gammaray dominance due to light-travel-time effects. In order to test this hypothesis and search for possible correlations with a time lag, we conduct the discrete correlation function analysis (DCF) \citep{edelsonkrolik1988}. The advantage of the DCF method is that it is particularly suitable for unevenly sampled data, which is exactly our case, and relies only on real measurements, without resorting to interpolations in the observational gaps. The DCF approach is widely used in astrophysics, and in blazar studies in particular, to explore possible correlations between multi-band light curves (e.g.\ \cite{kim2022}). We apply a numerical code implementing the DCF analysis to our light curves, to study correlations between different pairs of quantities, namely (1) emission line flux vs Compton Dominance, as well as (2) synchrotron continuum vs \gammaray flux, (3) emission line vs \gammaray flux and (4) synchrotron continuum vs emission line flux, in order to gain additional information about the complex connection between different observable parameters and establish which parameter(s) have the largest impact on the variations in another quantity. For each case, we calculate the DCF in the time lag values range -100 d $\leq \Delta t \leq$ 100 d (which slightly varies with the time lag binning), using a set of different values of time bin in the time lag space (5, 7, 10 and 14 days). The DCF as a function of time lag is shown in Fig.~\ref{fig:dcflfcdscgr} (dependence (1) to (4): from top to bottom).

From Fig.~\ref{fig:dcflfcdscgr} (top), one can see that the emission line flux does not display any pronounced correlation with the Compton dominance at any time lag value. At the same time, the optical synchrotron continuum shows a rather strong correlation with the \gammaray flux around zero time lag, as well as one can notice a presence of a significant peak in the DCF showing up at the time lag of $\sim 90$ d. Next, one can see a presence of modest correlation between the emission line luminosity and \gammaray flux occurring at a zero time lag, while the seeming feature appearing at $\sim$ -100 d lag is likely to be a fluctuation as the DCF errorbars around that point are very large, and we therefore choose to disregard this feature. Finally, one does not observe any substantial correlation between the optical synchrotron continuum flux and the emission line flux.

\begin{figure*}[t]
\includegraphics[width=0.8\textwidth]{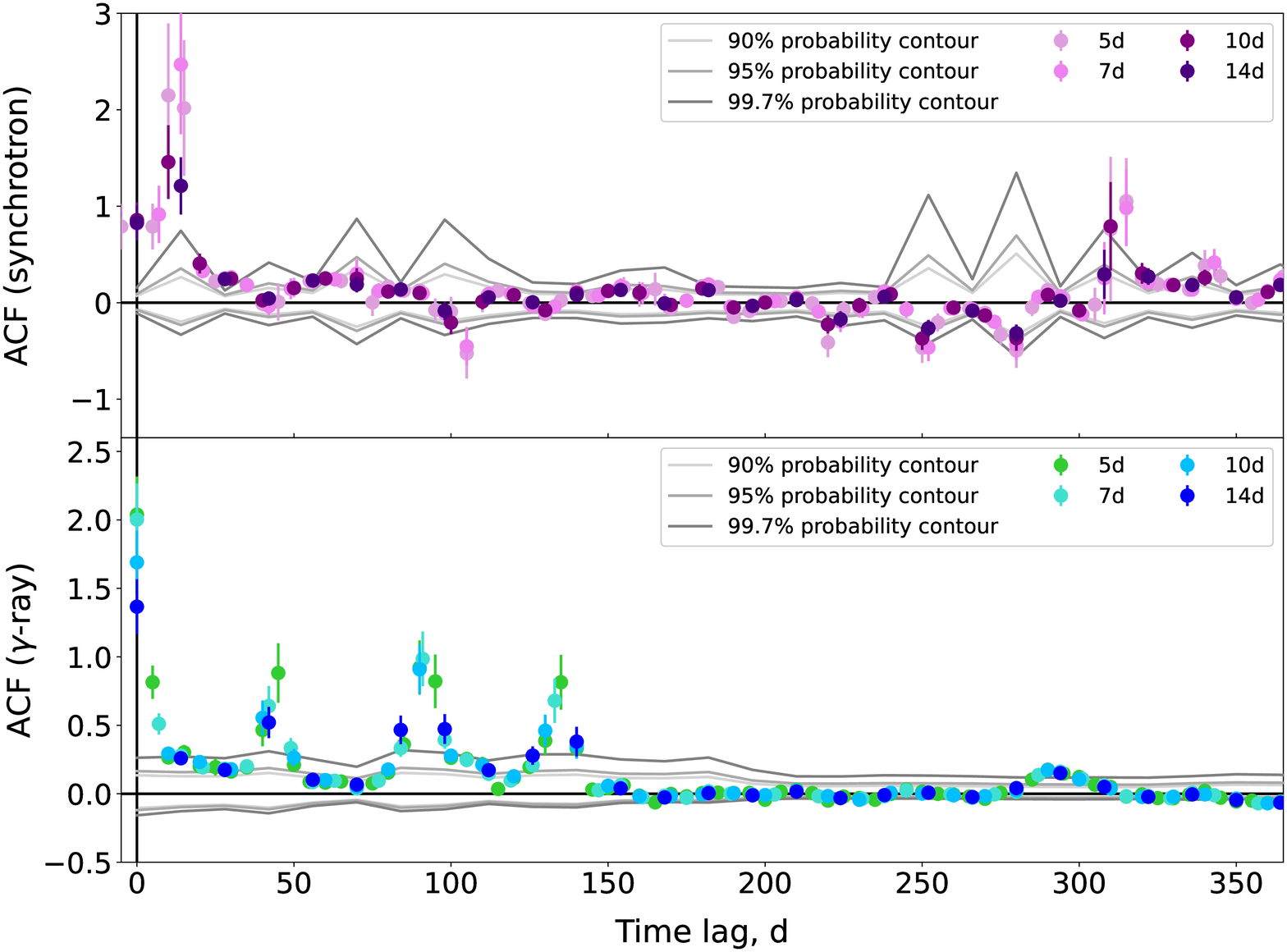}
\centering
\caption{ACF of the synchrotron continuum (top) and \Fermilat \gammaray (bottom) light curves. Different colors correspond to different binnings in the time lag space (5, 7, 10 and 14 day time bin widths). The gray contours, from lightest to darkest shades, represent 90\%, 95\% and 99.7\% ($3\sigma$) probability contours, respectively.}
\label{fig:acfsyngrf}
\end{figure*}

To further investigate the emergence of a moderate correlation between the optical synchrotron continuum and \gammaray light curves at a $\sim$ 90 d time lag, we perform the autocorrelation function (ACF) analysis for these light curves, which allows us to search for repetitive patterns in the long-term signal. The ACF is computed simply as the DCF with first and second input light curves being identical, in the time lag range spanning from 0 to 365 d. The resulting ACFs are shown in Fig.~\ref{fig:acfsyngrf} (synchrotron continuum flux -- top panel, \gammaray flux -- bottom panel). One can see that the ACF of the synchrotron continuum is essentially oscillating around zero and does not show any significant features. On the other hand, the ACF of the \gammaray light curve displays three distinct and statistically significant peaks, at $\sim 45$ d, $\sim 90$ d, and $\sim 135$ d. These time lag values appear to be first, second and third harmonic of a $\sim 45$ d period.

\subsection{Estimation of correlation significance} \label{sec:correlsignif}

In order to estimate the significance of the observed correlations from the DCF analysis, we perform Monte-Carlo (MC) simulations. We calculate a set of DCFs as a function of time lag, using the light curve of the emission line flux and of the Compton dominance, however, at each run randomly shuffling the data points of one of the light curves (the Compton dominance one was chosen arbitrarily). Qualitatively, we aim at estimating the highest DCF values that may arise by chance at a given probability level from the flux distributions defined by the given data sets. A total number of 30000 runs was performed, and the resulting DCFs were stacked. For this specific task, we have neglected the uncertainties on the DCF values. For each time lag, we determine the intervals containing 90\%, 95\% and 99.7\% of the realizations with positive DCF values, and the same intervals for negative DCF values, obtaining thus 90\%, 95\% and 99.7\% probability contour lines as a function of time lag. The 99.7\% contour indicates the 3$\sigma$ significance of the correlation. For the time lag binning, we use a 14-day bin which is considered as the optimal one, as for very small bin size one would have a large number of intervals lacking data points in the DCF calculation, leading to scarce statistics and large fluctuations in the contour values from bin to bin. We repeat the same procedure for the other three cross-correlations examined in the previous sub-section, as well as for the ACFs of synchrotron continuum and \gammaray light curves. The resulting probability contour lines are overlaid on top of the initial DCFs/ACFs, and are displayed in Fig.~\ref{fig:dcflfcdscgr} and Fig.~\ref{fig:acfsyngrf}.

\section{Discussion} \label{sec:discussion}

\subsection{Correlation between the optical synchrotron continuum and \gammaray flux}

\subsubsection{Two standing shocks scenario: \gammaray ``echo''}

The observed strong ($\sim 0.6 \pm 0.1$) correlation between the optical continuum and \gammaray flux at a zero time lag appears to be statistically significant ($> 3\sigma$) and thus indicates that those emissions are produced co-spatially (within a region of no more than a few light days, given the uncertainty on the DCF peak position), and by the same population of high-energy electrons, therefore favoring the leptonic scenario for 3C\,279. However, it is more challenging to explain the second peak in the DCF with the correlation degree $\sim 0.5 \pm 0.1$ (also significant above $3\sigma$), with the \gammaray emission variations lagging behind the ones in the optical band by $\sim 90$ d. Assuming a Doppler factor of $\delta \sim \Gamma \sim 10$, the time lag of $\Delta t \sim$ 90 d corresponds to a travel distance of $d \sim c \Delta t \Gamma \delta \sim 2.3 \times 10^{19}$ cm $\sim 7.5$ pc. Given the evidence of \gammaray generation at multi-parsec distances and even as far as $\sim 14$ pc from the central engine in BL Lac objects \citep[e.g.][]{agudo2011}, and assuming this might be the case in FSRQ objects as well, one can consider a scenario, in which a ``blob'' filled with electron-positron plasma and moving relativistically down the jet is crossing two standing shocks separated by the distance of $\sim$ 7 -- 8 pc. The first shock is located relatively close to the central engine (close to the BLR outer boundary assuming the IC-BLR scenario), and once the blob is passing through this shock, simultaneous optical synchrotron and \gammaray flares are produced, yielding the observed correlation at zero time lag. Then, once the blob travels $\sim$ 7 -- 8 pc along the jet and traverses the second shock, electrons in the blob are re-accelerated to high energies and, as the infrared emission from the dusty torus dominates the external radiation field at multi-parsec distances from the central engine, these electrons upscatter the infrared photons of the dusty torus to $\gamma$-rays via the inverse Compton process. Assuming the magnetic field in the jet at multi-parsec distances is very low, the synchrotron emission can be considered as negligible, and thus the initial simultaneous synchrotron and \gammaray flares will be accompanied by a \gammaray ``echo'' after a $\sim 90$ d delay. This will produce the observed second peak in the DCF at the $\sim 90$ d time lag. To verify this scenario in a quantitative manner, we perform a number of estimates of the conditions when such \gammaray flares can be prominent with a subdominant optical counterpart, given the relatively far distance from the central engine.

In order to produce a noticeable \gammaray flaring ``echo'', two conditions have to be fulfilled: (1) $U^{\prime}_{\mathrm{rad}} / U^{\prime}_B \gg 1$ for the synchrotron counterpart being negligible, and (2) $L_{\gamma,\mathrm{fl2}} \gtrsim L_{\gamma,\mathrm{q}}$ for the \gammaray flaring ``echo'' to be prominent above the quiescent emission level, with $L_{\gamma,\mathrm{fl2/q}}$ being \gammaray luminosities of the second (``echo'') flare and quiescent emission, respectively.

For our estimates, we are going to use emission modeling results from \cite{hayashida2012} developed for 3C\,279. This study is particularly suitable for our case since the authors consider a blob propagation along multi-parsec distances along the jet as one of the scenarios to explain data of a \gammaray flare. Here we adopt a two-zone model, with a larger unperturbed quiescent emitting region with a constant radius $R_{\mathrm{q}}$, and a smaller flaring blob having a constant radius $R_{\mathrm{fl}}$, moving along the jet with a bulk Lorentz factor $\Gamma \ = 10$ and interacting with two standing shocks. We neglect the particle escape here. Let us assume that the first shock crossing occurs at a distance of $r_1 = 1$ pc away from the central engine, then the second crossing happens at $r_2 = (1 + 7.5) \, {\rm pc} = 8.5$ pc. The energy densities of the BLR and dusty torus radiation fields (in the comoving frame) $U^{\prime}_{\mathrm{BLR/DT}}$ as a function of the distance $r$ from the central engine can be approximated as \citep{hayashida2012}: 

\begin{equation} \label{eq:u_rad_ext_r}
    U^{\prime}_{\mathrm{BLR/DT}} = \frac{\xi_{\mathrm{BLR/DT}} \, \Gamma^2 L_{\mathrm{D}}}{3 \pi r_{\mathrm{BLR/DT}}^2 \,  c \, [1 + (r/r_{\mathrm{BLR/DT}})^{\beta_{\mathrm{BLR/DT}}}]},
\end{equation}

with $\xi_{\mathrm{BLR/DT}}$ being the fraction of emission scattered by the BLR / dusty torus, $L_{\mathrm{D}} = 2 \times 10^{45}$ erg/s, $r_{\mathrm{BLR}} = 0.1 (L_{\mathrm{D}} / 10^{46} \text{ erg/s})^{1/2}$ pc, $r_{\mathrm{DT}} = 2.5 (L_{\mathrm{D}} / 10^{46} \text{ erg/s})^{1/2}$ pc, are the characteristic sizes of BLR and dusty torus, respectively, $\beta_{\mathrm{BLR}} = 3$ and $\beta_{\mathrm{DT}} = 4$ are indicies describing de-boosting of the BLR / dusty torus radiation fields with distance $r$ beyond their characteristic sizes. We adopt various parameter values from \cite{hayashida2012}, expect for scattering fractions and radius of the emitting zone, the values of which we (moderately) rescaled with respect to \cite{hayashida2012} to match the model values of $U^{\prime}_{\mathrm{BLR/DT}}$ and the \gammaray luminosity level, as the authors use slightly different bulk Lorentz factor values. Therefore, we use $\xi_{\mathrm{BLR/DT}} = 0.23$ $R_{\mathrm{fl}} = 3.7 \times 10^{17}$ cm (corresponding to the variability time-scale of $t_{\mathrm{var}} = 2$ weeks). Using Eq.~\ref{eq:u_rad_ext_r} we obtain the energy densities of the radiation fields at $r_1 = 1$ pc: $U^{\prime}_{\mathrm{BLR,1}} = 7.5 \times 10^{-4}$ erg/cm$^3$ and $U^{\prime}_{\mathrm{DT,1}} = 8.2 \times 10^{-4}$ erg/cm$^3$, and at $r_2 = 8.5$ pc: $U^{\prime}_{\mathrm{BLR,2}} = 1.2 \times 10^{-6}$ erg/cm$^3$ and $U^{\prime}_{\mathrm{DT,2}} = 4 \times 10^{-6}$ erg/cm$^3$. We estimate the magnetic field at the first shock crossing assuming a Compton dominance during the flare of $CD \sim 10$, which, using the relation $CD = (U^{\prime}_{\mathrm{BLR}} + U^{\prime}_{\mathrm{DT}})/U^{\prime}_B$ yields $B^{\prime}(r=1 \text{ pc}) \sim 0.15$ G, consistent with the model value from \cite{hayashida2012}. Next, we estimate the magnetic field at the second shock crossing considering two commonly used profiles of magnetic field decrease along the jet: (1) $B^{\prime} \propto 1/r$ \citep[e.g.][]{hayashida2012}, and (2) $B^{\prime} \propto 1/r^2$ \citep[e.g.][]{kaiser2006}. The resulting values of the magnetic field energy density at the distance $r_2 = 8.5$ pc for both profiles are: $U^{\prime}_B(r = 8.5 \text{ pc},1/r) = 1.2 \times 10^{-5}$ erg/cm$^3$ and $U^{\prime}_B(r = 8.5 \text{ pc},1/r^2) = 1.7 \times 10^{-7}$ erg/cm$^3$. One can see that, in order for the synchrotron counterpart of the flare to be negligible, the magnetic field along the jet should decrease as $\propto 1/r^2$. We therefore adhere to this profile here.     

Let us now examine the conditions for the \gammaray flaring emission produced at the second shock being above the quiescent level. We are specifically interested in the emission around $\sim 1$ GeV (in the observer's frame), since it falls in the \Fermilat energy range, and is also typically the energy at which the high-energy SED component peaks. The electrons in the blob interact with the external photon fields of BLR and dusty torus and generate a broad-band inverse Compton emission having two components. Approximating both photon fields as sharply peaked around $\epsilon^{\prime}_{\mathrm{BLR}} \sim 10 \times \Gamma$ eV and $\epsilon^{\prime}_{\mathrm{DT}} \sim 0.2 \times \Gamma$ eV (in the jet comoving frame), we obtain the Lorentz factors of electrons responsible for production of 1 GeV photons ($E^{\prime}_{\gamma} = 1 / \Gamma$ GeV): $\gamma_{\mathrm{BLR/DT}} = \sqrt{E^{\prime}_{\gamma} / \epsilon^{\prime}_{\mathrm{BLR/DT}}}$, $\gamma_{\mathrm{BLR}} = 10^{3}$ and $\gamma_{\mathrm{DT}} = 7 \times 10^3$. The exact shape of the electron spectrum (i.e.\ number of electrons at those Lorentz factors) can greatly affect the luminosities of the \gammaray emission components. 
The total \gammaray luminosity (during the quiescent state (``q'') or first/second flare (``fl1''/``fl2'')) around 1 GeV (in the observer's frame) is then:

\begin{multline} \label{eq:gammaraylumin}
    L_{\mathrm{\gamma,q/fl1/fl2}} \propto \left[N_{\mathrm{q / fl1 / fl2}}(\gamma_{\mathrm{BLR}}) U^{\prime}_{\mathrm{BLR,q/1/2}}\right. \\ + \left. N_{\mathrm{q / fl1 / fl2}}(\gamma_{\mathrm{DT}}) U^{\prime}_{\mathrm{DT,q/1/2}}\right] R_{\mathrm{q/fl}}^3 \Gamma^4 
\end{multline}

where $N_{\mathrm{q / fl1 / fl2}}(\gamma)$ is the electron spectrum in the quiescent (``q'') or flaring (``fl1/fl2'') states. We adopt the model of the quiescent emission from \cite{hayashida2012}, only (slightly) rescaling the radius of the emitting region to $R_{\mathrm{q}} = 1.5 \times 10^{18}$ cm ($t_{\mathrm{var}} \sim 2$ months) and the distance from the central engine to $r_{\mathrm{q}} = 3.4$ pc (due to a somewhat different Lorentz factor used in their model), in order to match again the total model external photon field energy density and the luminosity (and thus retaining the same electron distribution shape and normalization). The target BLR and dusty torus photon fields for the quiescent state are obtained via Eq.~\ref{eq:u_rad_ext_r}: $U^{\prime}_{\mathrm{BLR,q}} = 2 \times 10^{-5}$ erg/cm$^3$ and $U^{\prime}_{\mathrm{DT,q}} = 1.6 \times 10^{-4}$ erg/cm$^3$. The electron spectrum of the quiescent blob is represented as a broken power law with a break at $\gamma_{\mathrm{br,q}} = 400$ and with spectral indicies $\alpha_{\mathrm{q,1}} = 2.2$ and $\alpha_{\mathrm{q,2}} = 3.4$ below and above the break, respectively \citep{hayashida2012}. Next, for the flaring states, we have more freedom to choose an electron distribution form, and assuming acceleration of a ``cold'' / non-relativistic particle population residing in the flaring blob at a relativistic shock front (without injection) during the shock crossing, we adopt a simple power law with an index $\alpha_{\mathrm{fl}} = 2$. However, it is important to keep in mind that the accelerated particle population also undergoes strong radiative cooling, which leads to the formation of a cooling break at a Lorentz factor $\gamma_{\mathrm{c}} \sim (3 m_e c^2)/(2 \sigma_{\mathrm{T}} R_{\mathrm{fl}} \left[U^{\prime}_{\mathrm{BLR}} + U^{\prime}_{\mathrm{DT}} + U^{\prime}_B]\right)$. At the first shock crossing, the cooling break appears at $\gamma_{\mathrm{c,fl1}} \approx 510$, while at the second shock, the cooling break is at $\gamma_{\mathrm{c,fl2}} \approx 9 \times 10^5 \gg \gamma_{\mathrm{DT}}$. The generic form of the quiescent and flaring electron spectra is therefore $N_{\mathrm{q/fl1/fl2}} = A_{\mathrm{q/fl}} \gamma^{-\alpha_{\mathrm{q,1/fl,1}}}$ for $\gamma \leq \gamma_{\mathrm{break}}$, and $N_{\mathrm{q/fl1/fl2}} = A_{\mathrm{q/fl}} \gamma_{\mathrm{break}}^{\Delta \alpha} \gamma^{-\alpha_{\mathrm{q,2/fl,2}}}$ for $\gamma > \gamma_{\mathrm{break}}$, where for brevity we denote $\gamma_{\mathrm{break}} = \gamma_{\mathrm{br,q/c,fl1/c,fl2}}$ and $\Delta \alpha = \alpha_{\mathrm{q,2/fl,2}} - \alpha_{\mathrm{q,1/fl,1}}$. Knowing the parametrizations of the electron spectra, we can now estimate the ratio of the \gammaray luminosity (around 1 GeV) during the second shock crossing with respect to the quiescent level (using Eq.~\ref{eq:gammaraylumin}):

\begin{multline} \label{eq:ratiofl2toqgammaray}
    \frac{L_{\mathrm{\gamma,fl2}}}{L_{\mathrm{\gamma,q}}} = \frac{N_{\mathrm{fl2}}(\gamma_{\mathrm{BLR}}) U^{\prime}_{\mathrm{BLR,2}} + N_{\mathrm{fl2}}(\gamma_{\mathrm{DT}}) U^{\prime}_{\mathrm{DT,2}}}{N_{\mathrm{q}}(\gamma_{\mathrm{BLR}}) U^{\prime}_{\mathrm{BLR,q}} + N_{\mathrm{q}}(\gamma_{\mathrm{DT}}) U^{\prime}_{\mathrm{DT,q}}} \left(\frac{R_{\mathrm{fl}}}{R_{\mathrm{q}}}\right)^3 \\ = \frac{A_{\mathrm{fl}}}{A_{\mathrm{q}}}  \, \gamma_{\mathrm{br,q}}^{\alpha_{\mathrm{q,1}} - \alpha_{\mathrm{q,2}}} \, \frac{\gamma_{\mathrm{BLR}}^{-\alpha_{\mathrm{fl,1}}} U^{\prime}_{\mathrm{BLR,2}} + \gamma_{\mathrm{DT}}^{-\alpha_{\mathrm{fl,1}}} U^{\prime}_{\mathrm{DT,2}}}{\gamma_{\mathrm{BLR}}^{-\alpha_{\mathrm{q,2}}} U^{\prime}_{\mathrm{BLR,q}} + \gamma_{\mathrm{DT}}^{-\alpha_{\mathrm{q,2}}} U^{\prime}_{\mathrm{DT,q}}} \left(\frac{R_{\mathrm{fl}}}{R_{\mathrm{q}}}\right)^3
\end{multline}

Imposing a modest requirement $L_{\mathrm{\gamma,fl2}} / L_{\mathrm{\gamma,q}} \simeq 2$, we retrieve a condition $A_{\mathrm{fl}} / A_{\mathrm{q}} \approx 169$, which quantifies the requirement of a sufficient number of particles in the population $N_{\mathrm{fl2}}(\gamma)$ to produce GeV \gammaray emission at a distance of 8.5 pc that is prominent above the quiescent level. We consider that before crossing the shock, particles in the flaring blob are cold, non-relativistic, and do not radiate non-thermal emission. During the shock acceleration, particles migrate to higher energies, and after the shock crossing the electron spectrum attains the shape of $N_{\mathrm{fl1/fl2}}(\gamma)$. Assuming that both quiescent and flaring blobs have similar levels of (total) particle number densities (owing, for instance, to the same injection mechanism which we do not consider here), and without detailed modeling of the electron spectrum evolution during the acceleration on the shock front, we verify whether the total number of cold particles in the flaring blob is sufficient to give rise to the electron spectrum $N_{\mathrm{fl2}}(\gamma)$ with the normalization constrain derived above. The number density of the quiescent electron spectrum is $n_{\mathrm{q}} \simeq \int_{\gamma_{\mathrm{min,q}}}^{\gamma_{\mathrm{br,q}}} N_{\mathrm{q}}(\gamma) d\gamma \simeq A_{\mathrm{q}} (\alpha_{\mathrm{q,1}} - 1)^{-1} \left[\gamma_{\mathrm{min,q}}^{1 - \alpha_{\mathrm{q,1}}} - \gamma_{\mathrm{br,q}}^{1 - \alpha_{\mathrm{q,1}}} \right]$, where $\gamma_{\mathrm{min,q}}$ is the minimum Lorentz factor of the $N_{\mathrm{q}}$ spectrum. The electron number density in the flaring blob, following the same approach, is $n_{\mathrm{fl}} \simeq A_{\mathrm{fl}} / \gamma_{\mathrm{min,fl}}$, with $\gamma_{\mathrm{min,fl}}$ being the minimum Lorentz factor of the $N_{\mathrm{fl1/fl2}}$ spectrum. As we assume that $n_{\mathrm{q}} \simeq n_{\mathrm{fl}}$, and requiring $\gamma_{\mathrm{min,fl}} \leq \gamma_{\mathrm{BLR}}$ (i.e.\ requiring electrons to produce IC-BLR emission), we obtain that $\gamma_{\mathrm{min,q}}$ should be $\sim 4$ (or less). Although the modeling by \cite{hayashida2012} does not specify the value of $\gamma_{\mathrm{min,q}}$, this condition however can be relaxed if we assume that the shock injects particles into the flaring blob during the crossing. It is rather difficult to accurately quantify such injection without a detailed modeling, but we suppose here that after the second shock crossing, the total number of particles in the flaring blob is similar to the one in the quiescent blob, i.e.\ $n_{\mathrm{q}} R_{\mathrm{q}}^3 \simeq n_{\mathrm{fl}} R_{\mathrm{fl}}^3$. This relation yields a less stringent condition that $\gamma_{\mathrm{min,q}}$ has to be $\sim 100$ (or less), which appears to be reasonable. Therefore, there should be enough particles in the flaring blob to produce strong \gammaray emission at distances of 8.5 pc during a shock crossing that is depositing particles into the blob.

Another question is whether the \gammaray flaring emissions produced at the two shocks have drastically different luminosities (and so whether the flare at the first shock has unrealistically high luminosity and flux increase factor above the quiescent level), since the radiation fields densities differ by around three orders of magnitude. It is worth to note, however, that there is no strict requirement for the luminosity ratio of the two flares based on ACF/DCF results. In fact, the \gammaray ``echo'' can, in principle, achieve a luminosity comparable to the one at the first shock crossing (or at least satisfy $L_{\mathrm{\gamma,fl1}} / L_{\mathrm{\gamma,fl2}} \ll 10^3$), due to (1) $N_{\mathrm{fl2}}(\gamma)$ having a much higher cooling break ($\gamma_{\mathrm{c,fl2}} \approx 9 \times 10^5$) compared to $N_{\mathrm{fl1}}(\gamma)$ ($\gamma_{\mathrm{c,fl1}} \approx 510$) and therefore much more electrons available around $\gamma_{\mathrm{BLR}} = 10^3$ and $\gamma_{\mathrm{DT}} = 7 \times 10^3$ for upscattering photon fields, and (2) the possible injection of additional particles at the second shock. One can attempt to quantify this effect using Eq.~\ref{eq:gammaraylumin}. However, the exact ratio of luminosities strongly depends on the input parameters, namely (1) the exact location of the cooling break, which depends on the exact spectra of the BLR and dusty torus photon fields, as well as the microphysics of the shock acceleration process, and (2) the number of particles deposited by the second shock into the blob. Depending on the (reasonable) choice of these quantities, the ratio between the \gammaray luminosities of the two flares can be as low as $\sim 10$. 
  
Therefore, within such a scenario it indeed seems to be possible to produce a non-negligible \gammaray ``echo'' under certain (reasonable) physical conditions. Obviously, detailed numerical simulations using radiative codes are required to investigate this model more thoroughly.

\subsubsection{Multiple recollimation shocks scenario: repetitive \gammaray signal}

The ACF analysis performed for the synchrotron continuum and \gammaray flux can provide additional clues about the origin of the above-mentioned $\sim 90$ d peak in the DCF. This time lag corresponds to the second harmonic in the ACF of the \Fermilat light curve (Fig.~\ref{fig:acfsyngrf}, bottom panel). Therefore, it is possible that the $\sim 90$ d period in the optical synchrotron versus \gammaray DCF is due to a repetitive pattern in the \gammaray signal. The $\sim 45$ d period observed in the ACF of the $\gamma$-ray light curve can, for instance, be explained within the multiple internal shock scenario \citep[e.g.][]{hervet2017}. In this picture, a relativistic jet with transverse stratification features a sequence of multiple recollimation shocks (possibly associated with knots observed in the radio jet structure). Electrons interacting with the successive shocks undergo acceleration and produce flaring emission with a repetitive pattern \citep[e.g.][]{hervet2019}. The period of $\sim 45$ d in the ACF then corresponds to a distance between two successive shocks of $\sim 1.2 \times 10^{19}$ cm $\sim 3.8$ pc. Given the presence of only three harmonics in the \gammaray light curve ACF, we can conclude that the jet of 3C\,279 has three recollimation shocks (knots) separated by an equal distance of $\sim 3.8$ pc. However, it is unclear why the first harmonic ($\sim 45$ d time lag) does not manifest in the DCF of the optical synchrotron versus $\gamma$-rays, while there is a hint of a (marginally significant) peak in the DCF of emission lines versus $\gamma$-rays. A more detailed analysis, which is beyond the scope of this paper, would be necessary to understand these details.

\subsection{Correlation between the optical synchrotron continuum and emission line flux}

Next, the synchrotron continuum does not show any correlation with the emission line flux, which means that the varying synchrotron emission from the jet does not influence the generation of emission lines in the BLR. This suggests that the variations in the emission line luminosity most likely are arising due to the varying luminosity of the accretion disk emission, which is reprocessed by the BLR.

\subsection{Correlation between the emission line flux and \gammaray flux and Compton dominance}

Then, we identify a moderate positive correlation ($\sim 0.5 \pm 0.25$) between the emission line and \gammaray flux appearing at a zero time lag (or a slightly negative one), found to be marginally significant (at a level slightly higher than $3\sigma$). At the same time, no significant correlation is found between the the emission line flux and the Compton dominance variations. This has several possible interpretations.

A slight correlation between the emission line luminosity and \gammaray flux that we observe at a zero time lag can be explained by the (initially assumed) IC-BLR process. Indeed, within the IC-BLR scenario, the \gammaray luminosity observed by a distant observer is $L_{\gamma} \propto N_e U^{\prime}_{\mathrm{rad,BLR}} \delta^4 \propto N_e L_{\mathrm{eml}} \Gamma^2 \delta^4$, while the synchrotron continuum luminosity is $L_{\mathrm{c}} \propto N_e B^2 \delta^4$. One can see that the \gammaray luminosity is proportional to the emission line luminosity, but also appears to be dependent on other parameters, $N_e$, $\Gamma$ and $\delta$, the variations of which will break the proportionality relation. Thus, a clear correlation between the emission line luminosity and \gammaray flux will only be observed if the relative variations in the electron number density and the bulk Lorentz factor of the jet (more precisely, the quantity $\Gamma^2 \delta^4$) remain weaker than (or comparable to) the ones in the emission line luminosity. Our results therefore seem to indicate approximately equal relative amplitudes in variations of these parameters.

The Compton dominance is $CD = L_{\gamma} / L_{\mathrm{c}} \propto L_{\mathrm{eml}} \Gamma^2 / B^2$. Again, a clear linear correlation between the Compton dominance and the emission line flux will only hold in case the magnetic field and the Lorentz factor of the emitting zone remain constant (or vary much more weakly than the emission line luminosity). Therefore, the non-observation of a correlation between the emission line flux and the Compton dominance might arise from the fact that the Compton dominance changes are being mostly caused by either the magnetic field variations or the ones in the bulk Lorentz factor. As previously we established that the emission line luminosity and the quantity $\Gamma^2 \delta^4$ should have commensurate relative amplitudes of variability, the quantity $\Gamma^2$ should provide an impact on the Compton dominance similar to (or less than) the one of the emission lines. Thus, the lack of a clear correlation between emission-line luminosity and CD suggests that magnetic field changes have the strongest relative magnitude and represent the dominant parameter driving the dramatic variations in the Compton dominance, suppressing any contribution from emission line luminosity or the Lorentz factor variations.

Unfortunately, it remains quite difficult to isolate the effect of one specific parameter on another one and infer to which quantity are variations in the \gammaray flux and Compton dominance are connected. This might be the reason why a number of previous optical-\gammaray correlation studies show rather unclear results. For example, in a study by \cite{rajput2020}, the authors identify three types of relations between the optical and \gammaray variations in 3C\,279 (as well as a number of other blazars), namely (\textit{a}) a pronounced correlation, (\textit{b}) \gammaray flares without an optical counterpart (``orphan'' flares), and (\textit{c}) optical flares without a \gammaray flux increase (``sterile'' flares). Based on the SED study performed by the authors, the case (\textit{a}) is attributed to variations in the bulk Lorentz factor, the case (\textit{b}) to changes in the electron energy density, and the case (\textit{c}) to an amplification of the magnetic field in the emitting region.

Another interband correlation analysis by \cite{larionov2020} is based on a 10-year MWL data set of 3C\,279 including optical measurements from the GASP–WEBT programme, the \Fermilat telescope, and other instruments. First, the study reveals a moderate correlation between the optical and \gammaray flux of the order of DCF $\sim 0.4$ with the $\gamma$-rays leading the optical by $1.06 \pm 0.47$ d, which is overall compatible with our result on the optical versus \gammaray flux correlation. By directly comparing the $R$-band and \Fermilat light curves, the authors find orphan and sterile flares, and discover that the optical-\gammaray relationship depends on the activity state of the object. Four time intervals are identified, with slopes of $\text{log}(F_{\gamma}/F_R)$ being (1) $0.97 \pm 0.04$ during 2008-2010, then a sharp change to (2) $7.7 \pm 1.2$ during 2011-2016, (3) $0.82 \pm 0.06$ during a short period in the end of 2016 characterized by a violent optical flare, and (4) $1.9 \pm 0.09$ after the flare peak (early 2017 -- 2018). The authors attribute the measured slopes to different physical scenarios. The slope (1) is interpreted as mainly due to an increase in the electron number density, with a contribution of changes in the Doppler factor due to jet bending and/or bulk Lorentz factor variations. The steep slope (2) is considered to be appearing due to both variations in the number of radiating electrons and in the energy density of seed photons, with the changes in the latter tentatively explained as being triggered by synchrotron flares, as a correlation between the continuum and emission line flux with a time delay less than 2 months is revealed in the analysis by the authors. To the contrary, in our analysis we do not observe any significant correlation between these two quantities. The slope (3) is described by a complex scenario with a combination of magnetic field growth and decrease in the seed photon energy density, possibly as well accompanied by variations in the electron number density. Finally, the slope (4) is ascribed to a superposition of changes in the seed photon energy density and the in the bulk Lorentz factor. These results clearly demonstrate that the flux variations in both synchrotron continuum and \gammaray band are caused by a complex combination of different physical processes dependent on the activity state of the source, and disentangling those processes proves to be a rather challenging task.

One can attempt to use various combinations of the directly observable quantities, namely $L_{\gamma}$, $L_{\mathrm{c}}$ and $L_{\mathrm{eml}}$, to exclude the influence of the (in our case) non-measurable variables $B$, $\Gamma$, and $N_e$. However, degeneracies remain, even assuming an equipartition relation, such as $N_e \propto B^2$ \citep[e.g.,][]{dermer2014}. In calculating the Compton dominance, the electron number evolution is eliminated, while the \gammaray light curve variations are independent of the magnetic field changes. To exclude the dependence on the Lorentz factor, one can examine a parameter $f_{\gamma-c-3/2} = L_{\gamma} / L_{\mathrm{c}}^{3/2} \propto N_e^{-1/2} B^{-3} L_{\mathrm{eml}}$, which is $\propto B^{-4} L_{\mathrm{eml}}$ assuming equipartition. Finally, with the same assumption, one can eliminate the magnetic field, $f_{\gamma-c-1/2} = L_{\gamma} / L_{\mathrm{c}}^{1/2} \propto L_{\mathrm{eml}} \Gamma^4$. The two last quantities allow us to explore the impact of the emission line luminosity on the \gammaray flux in the presence of only one disturbing parameter. If one observes a correlation of both $f_{\gamma-c-3/2}$ and $f_{\gamma-c-1/2}$ with the line flux at the same time lag, one could conclude that the emission line luminosity variations indeed influence those of the \gammaray luminosity. Having a measurement of the frequencies of the two SED peaks can help to lift the degeneracy and constrain the necessary parameters quite precisely. However, unfortunately, those data are not available with a regular sampling (and enough statistics).

Another explanation for the absence of a correlation between the emission line luminosity and the Compton dominance could be that the IC-BLR mechanism might not always be responsible for production of GeV \gammaray emission in 3C\,279, leading also to relatively weak correlation between the emission line flux and the \gammaray flux. For example, while the March-April 2014 \gammaray flare of the source is well explained in an external Compton scenario with both BLR and dusty torus photons, during low states of the source, the dusty torus might provide a dominant contribution to the GeV \gammaray production (\citep[e.g.][]{hayashida2012}). Finally, another option is that hadronic processes might manifest themselves at certain activity states of the object. For example, its December 2013 flare \citep{paliya2016}, or the minute-scale GeV flare in June 2015 \citep{petropoulou2017} both posed serious challenges to leptonic models, with the observed spectral and timing flux properties much more easily accommodated within a hadronic description. A detailed cross-correlation study of separate subsets of data for various stages of source activity is required to gain more insight into this problem.

Establishing the source of external target photons is important for constraining the location of the emitting zone within the jet. As already discussed, the IC-BLR hypothesis for 3C\,279 seems the most plausible given the set of observed correlations. Therefore, the \gammaray production site should be located in the vicinity of the BLR, but cannot reside too deep within it: \cite{paliya2015} and \cite{bottcherels} show that the emitting zone of 3C\,279 must be located close to the BLR outer boundary, rather than the inner one, so that one (1) avoids significant $\gamma$-$\gamma$ opacity, as well as (2) explains a non-negligible contribution of the dusty torus photon field to the \gammaray production required to explain the March-April 2014 flare of the source.

It is also worth mentioning that the quality of data used in this study (especially in the optical domain) might have impacted the degree of observed correlations: some of the optical spectra often displayed a noticeable level of noise, which leads to inaccuracies in the determination of the emission line flux. Another important factor is a possible selection bias, due to (1) uneven time sampling of the optical data (certain states of the source might have been observed more frequently), and/or (2) a lack of precise emission line measurements during high levels of optical synchrotron flux (when the continuum outshines the emission lines).

\section{Summary and Outlook} \label{sec:summary}

The main results of this work can be summarized as follows:

\begin{enumerate}
\item Based on a 10-year optical and \gammaray data set of 3C\,279, we identify a pronounced correlation ($\sim 0.6 \pm 0.1$) between the optical synchrotron continuum flux and the GeV \gammaray flux occurring at a time lag consistent with zero. This means that these two emissions are produced co-spatially by the same electron population and supports a leptonic \gammaray emission origin. 
\item In addition, we find that the synchrotron continuum and \gammaray flux display also a moderate correlation ($\sim 0.5 \pm 0.1$) at a time lag $\sim 90$ d, with the $\gamma$-ray flux variations lagging behind the ones of the synchrotron flux. We interpret this result in a scenario where a blob filled with electron-positron plasma and traveling along the jet axis is traversing two standing shocks separated by a distance of $\sim$ 7 -- 8 pc, corresponding to the 90 d time lag. Once the blob is crossing the second shock, re-accelerated electrons generate \gammaray emission via inverse Compton scattering of dusty torus photons, leading to the \gammaray ``echo'' which follows the flare emission from the crossing of the first shock after a 90-day delay.
\item No substantial correlation is seen between the optical synchrotron continuum flux and the emission line luminosity, which implies that amplifications of the BLR emission line fluxes are mostly triggered by high states of the accretion disk, rather than the optical synchrotron flares originating in the jet.
\item No significant correlation is observed between the emission line luminosity and the Compton dominance.
\item We also reveal a moderate, marginally significant correlation ($\sim 0.5 \pm 0.25$) between the emission line luminosity and the \gammaray flux appearing at a zero time lag, which points to an IC-BLR GeV \gammaray emission origin, with the \gammaray production site located close to the BLR outer boundary (to avoid the $\gamma$-$\gamma$ opacity). The observed correlation also indicates that the variations of the electron number density and the bulk Lorentz factor (more precisely, the quantity $\Gamma^2 \delta^4$) cannot be substantially stronger than the ones of the emission line luminosity. Therefore, one can infer roughly equal contributions from these three parameters to the \gammaray flux variability.
\item The fact that the emission line luminosity displays a correlation with the \gammaray flux, but not with the Compton dominance, suggests that changes in the Compton dominance are mostly driven by strong magnetic field variations, with the emission line luminosity and bulk Lorentz factor variations being much weaker and playing a minor role.
\end{enumerate}

\section{Acknowledgements}
The work of M.B. was supported by the South African Research Chairs Initiative of the National Research Foundation\footnote{Any opinion,
finding, and conclusion or recommendation expressed in this material is that of the authors, and the NRF does not accept any liability in this regard.} and the Department of Science and Innovation of South Africa through SARChI grant no. 64789. A.D. is grateful to Michael Zacharias and Catherine Boisson for fruitful discussions regarding this study, which allowed to significantly improve the results. Data from the Steward Observatory spectropolarimetric monitoring project were used. This program is supported by Fermi Guest Investigator grants NNX08AW56G, NNX09AU10G, NNX12AO93G, and NNX15AU81G.

\bibliography{3c279}{}
\bibliographystyle{aasjournal}



\end{document}